\documentstyle[preprint,aps]{revtex}

\begin{document}
\topmargin -1.4cm

\draft
\title{Correlated two-electron transport: a principle for a novel
charge pump}

\author{Qing-feng Sun$^1$, Hong Guo$^1$, and Jian Wang$^2$}

\address{$^1$Center for the Physics of Materials and Department
of Physics, McGill University, Montreal, PQ, Canada H3A 2T8.\\
$^2$Department of Physics, The University of Hong Kong, Pokfulam Rood,
Hong Kong, China}

\maketitle

\begin{abstract}
By considering a correlated two-electron transport process (TET) and using
a diagrammatic analysis within the Keldysh nonequilibrium Green's function 
formalism, we discuss a novel charge pump by which carriers are pumped from 
a contact with low chemical potential to another contact with a higher 
potential. The TET process involves two correlated incident electrons 
scattering and exchanging energy with each other. The process can significantly 
affect charge current density and it involves high empty states and/or low 
filled states of the Fermi liquid of the leads. 
\end{abstract}

In the past two decades, quantum electron transport through meso- and 
nanoscopoic structures have received considerable attention\cite{datta_book}.  
Charge transport theories for nanostructures typically consider the situation 
of an incident electron from a lead that is scattered in the device scattering 
region and transmitted to other leads or reflected to where it came from. 
The scattering region can be rather complicated and may involve 
electron-electron, electron-phonon, electron-impurity, and other interactions 
and scattering mechanisms. The scattering region itself can be a semiconductor 
nanostructure such as a quantum dot (QD), a carbon nanotube, or a 
single molecule. The scattering processes may involve exchange of energy or 
spin. While situations and physics can vary in a wide range of ways, charge 
transport in nanostructures has largely, so far, been considered as involving 
a single electron incident from a non-interacting lead of the device,
it traverses through a complicated scattering region in which interactions 
occur, and finally exits the device through the non-interacting leads. We will 
refer them as single electron tunneling or single electron transport (SET) 
processes.

In this paper, we go one step further by investigating {\it correlated} 
two-electron tunneling and transport (TET) processes which, to the best of 
our knowledge, has not been studied before. In this process, two incident 
electrons with energies $\epsilon_1$ and $\epsilon_2$ from the leads tunnel 
into the scattering region of a device, they scatter with each other and 
exchange energy (they could also scatter and exchange energy with other 
particles in the scattering region). Following the scattering event, they 
transmit to the outside world through the contacts, but with energies 
$\epsilon_1^{'}$ and $\epsilon_2^{'}$ which are different from the initial 
values $\epsilon_1$, $\epsilon_2$ due to energy exchange. If 
$\epsilon_1+\epsilon_2 = \epsilon_1^{'} +\epsilon_2^{'}$, {\it i.e.} the two 
incident electrons do not exchange energy with other particles, the transport 
process as a whole can be considered as elastic. In the TET process, the 
variation of electron number in the scattering region is always {\it two}: a 
situation that is different from SET. We show that such TET processes
can indeed occur, and it affects measurable physical quantities such as 
charge current. In particular, TET induces a transport vertical 
flow---meaning carriers incident with energy $\epsilon_1 $ exit at a 
different energy $\epsilon_1'$, resulting to a non-conserved current 
{\it density}, {\it i.e.} $\sum_nj_n(\epsilon)\not=0$. Using the 
characteristics of the correlated two-electron transport process, we design 
an interesting charge pump so that incident carriers can be pumped from a 
contact with low chemical potential to another contact having a higher 
potential.

We consider a QD coupled to two or three leads, described by the following 
Hamiltonian:
$ H  = \sum_{\alpha} \epsilon_{\alpha}
d^{\dagger}_{\alpha}
 d_{\alpha} + Ud^{\dagger}_{\uparrow} d_{\uparrow}
 d^{\dagger}_{\downarrow} d_{\downarrow} 
+\sum_{n,k,\alpha}
 \epsilon_{nk} a^{\dagger}_{nk\alpha} a_{nk\alpha}
 +\sum_{n,k,\alpha} [t_{nk} a^{\dagger}_{nk\alpha} d_{\alpha}
    +H.c.]
$,
where $a^{\dagger}_{nk\alpha}$($a_{nk\alpha}$) and
$d^{\dagger}_{\alpha}$($d_{\alpha}$) are creation (annihilation) operators
in lead $n$ and in QD, respectively. The QD includes two states and has an 
intradot Coulomb interaction $U$. Subscript $\alpha$ is the spin index, it 
may also indicate other quantum numbers. To account for a magnetic field, 
we let $\epsilon_{\uparrow}\not= \epsilon_{\downarrow}$. 

In the following we focus on investigating the elastic TET by analyzing the
behavior of current density $j_n(\epsilon)$ from lead $n$ to the QD.
$j_n(\epsilon)$ relates to the current $I_n$ through 
$I_n=\int j_n(\epsilon)d \epsilon$. We also define the electron occupation 
number density operator
$\hat{N}_n(\epsilon,\tau) = \sum_{k,\alpha} \int e^{i\epsilon t}
a^{\dagger}_{nk\alpha}(\tau) a_{nk\alpha}(\tau +t) \frac{dt}{2\pi}$.
$\hat{N}_n(\epsilon,\tau) d\epsilon$ describes the electron occupation
number in lead $n$ in the energy range $\epsilon$ to $\epsilon+d\epsilon$ at
time $\tau$. Current density $j_n(\epsilon)$ can be calculated from the time
evolution of $\hat{N}_n(\epsilon,\tau)$:
$j_{n}(\epsilon)=-e<\frac{d}{d\tau} \hat{N}_n(\epsilon,\tau)>$. This leads to
(in units of $\hbar=1$):\cite{meir1}
\begin{equation}
  j_n(\epsilon) =-e Im \sum\limits_{\alpha}
    \frac{\Gamma_n(\epsilon)}{2\pi}
    \left[ 2f_n(\epsilon) G^r_{\alpha}(\epsilon) +G^<_{\alpha}(\epsilon)
    \right]
\end{equation}
where linewidth function $\Gamma_n(\epsilon) \equiv 2\pi \sum_k
|t_{nk}|^2 \delta(\epsilon-\epsilon_{nk})$; $f_n(\epsilon)$ is the Fermi
distribution function for lead $n$; $G^{r,<}_{\alpha}(\epsilon)$ are the 
retarded and Keldysh Green's functions of the QD.\cite{ref4} Using the
standard equation of motion technique, $G^r_{\alpha}(\epsilon)$ has already
been solved in previous work:\cite{ref4,ref5}
\begin{equation}
  G^r_{\alpha}(\epsilon) = \frac{1+UA_{\alpha} n_{\bar{\alpha}} }
    {\epsilon-\epsilon_{\alpha} -\Sigma^{(0)}_{\alpha}(\epsilon)
      +U A_{\alpha} (\Sigma^a_{\bar{\alpha}} +\Sigma^b_{\bar{\alpha}} )} 
\end{equation}
where $A_{\alpha} (\epsilon) = [\epsilon-\epsilon_{\alpha} -U 
-\Sigma^{(0)}_{\alpha}(\epsilon) -\Sigma^{(1)}_{\bar{\alpha}}(\epsilon)
-\Sigma^{(2)}_{\bar{\alpha}}(\epsilon) ]^{-1}$;
$\Sigma^{(0)}_{\alpha} $
is the lowest-order self-energy, $\Sigma_{\alpha}^{(1),(2),a,b}$ are 
the higher-order self-energies;\cite{ref4}
and $n_{\alpha} = 
Im \int \frac{d\epsilon}{2\pi} G^<_{\alpha}(\epsilon) $ is the
intradot electron occupation number of state $\alpha$.
It is worth to mention that if temperature $T$ is lower than the Kondo 
temperature $T_K$, the solution of Eq.(2) has a Kondo resonance at the 
Fermi level which was the subject of many previous studies.\cite{ref5} 

We investigate TET processes at a temperature higher than $T_K$. To this 
end we need to solve the the Keldysh Green's function $G^<_{\alpha}(\epsilon)$.
Note that if one applies the commonly used ansatz for interacting lesser and 
greater self-energies\cite{ref6}, or using the large-$U$ limit non-crossing 
approximation\cite{ref5,ref7} to solve $G^<_{\alpha}(\epsilon)$, the 
two-electron scattering will be lost in these approximations. Therefore a 
more precise analysis is needed in our problem and we proceed as follows. 
Introducing the intradot electron occupation number density operator
$\hat{N}_{\alpha}(\epsilon,\tau) =\int e^{i\epsilon t}
d^{\dagger}_{\alpha}(\tau) d_{\alpha}(\tau +t) \frac{dt}{2\pi} $
and due to the steady state condition 
$<\frac{d}{d\tau}\hat{N}_{\alpha}(\epsilon,\tau) > =0 $, we have:
\begin{equation}
  -e \sum\limits_{n} \frac{i\Gamma_{n}}{2\pi} \left[
   G^<_{\alpha} +f_n(G^r_{\alpha}-G^a_{\alpha}) \right] 
  = i_{\alpha}(\epsilon)
\end{equation}
where $i_{\alpha}(\epsilon)$ gives the intradot vertical flow:
$ i_{\alpha}(\epsilon) \equiv ieU \int e^{i\epsilon t}
  \left\{ <d^{\dagger}_{\alpha}(0) d^{\dagger}_{\bar{\alpha}}(0)
           d_{\bar{\alpha}}(0) d_{\alpha}(t) > 
    - <d^{\dagger}_{\alpha}(0) d^{\dagger}_{\bar{\alpha}}(t) 
           d_{\bar{\alpha}}(t) d_{\alpha}(t) >
  \right\} \frac{dt}{2\pi} $.
The quantity $i_{\alpha}(\epsilon)$ can be viewed as an intradot ``vertical'' 
current density\cite{datta_book}
at energy $\epsilon$ that is contributed by carriers with 
other energies due to the e-e scattering process. Summing over the
index $\alpha$, Eq.(3) reduces to $\sum_n j_n(\epsilon) +\sum_{\alpha}
i_{\alpha}(\epsilon) =0$, which is exactly the steady state current 
conservation equation so that the total current density---including the 
vertical flow, through the intradot energy level $\epsilon$ is zero. If 
$G^r_{\alpha}(\epsilon)$ and $i_{\alpha}(\epsilon)$ have been solved, from 
Eq.(3) $G^<_{\alpha}(\epsilon)$ can be obtained immediately.

Next, we solve the vertical flow quantity $i_{\alpha}(\epsilon)$. We introduce
two particle contour-ordered Green's functions $B_{\alpha}(t,0)$ and 
$B_{c}(t,0)$, $B_{\alpha}(t,0) \equiv -< T_C [d_{\alpha}(t) 
d_{\bar{\alpha}}(0^+) d^{\dagger}_{\alpha}(0) d^{\dagger}_{\bar{\alpha}}(0) ]>$
and $B_{c}(t,0) \equiv -< T_C [d_{\alpha}(t) d_{\bar{\alpha}}(t) 
d^{\dagger}_{\alpha}(0) d^{\dagger}_{\bar{\alpha}}(0) ]>$.
Although there are four operators in the definition $B_{\alpha/c}(t,0)$, 
only two time indices at $(t,0)$ appear. We can therefore write the contour 
ordered quantities as $B_{\alpha/c}^{++}(t,0)$, $B_{\alpha/c}^{+-}$,
$B_{\alpha/c}^{-+}$, and $B_{\alpha/c}^{--}$. Their Fourier transformations
can be defined as $B(\epsilon) \equiv \int e^{i\epsilon t} B(t,0) dt$.
Using the two-particle Green's function, the vertical flow
$i_{\alpha}(\epsilon)$ is reduced to
$i_{\alpha}(\epsilon) =-\frac{eU}{\pi} Im B_{\alpha}^{+-}(\epsilon)$.

The contour Green's function $B_{\alpha/c}$ is solved by means of a
Feynman diagram expansion using the Wick's theorem. We take the interacting 
part of the Hamiltonian as $H_I=Ud^{\dagger}_{\uparrow}d_{\uparrow}
d^{\dagger}_{\downarrow}d_{\downarrow}$, and the remaining part 
$H-H_I$ as the non-interacting Hamiltonian $H_0$. The first-order irreducible 
self-energy which we consider is shown in Fig.1. 
It is clear that 
the first-order graph describes a two-particle propagation 
involving the exchange of an interacting energy $U$. 
Two important points should be mentioned here. (i) The 
vertex we calculate sums up all reducible diagrams constructed by the 
irreducible self-energy of Fig.1 (also see Eq.(4)). This level of
approximation
is equivalent to that of a typical random phase approximation, {\it i.e.}
we compute the irreducible self-energy up to order $U^{-1}$.  It is reasonable
to neglect other higher order irreducible diagrams which are of orders 
$U^{-2}$ and higher, because the interaction energy $U$ is large. (ii)
Single solid lines in Fig.1 stand for the intradot contour-ordered Green's 
functions $G_{\alpha}$ of the Hamiltonian $H$ (not $H_0$). This means
that we have summed over all terms in the Feynman diagram expansion of
$G_{\alpha}$.

The corresponding equation for Fig.1 is
\begin{eqnarray}
& &  B_{c/\alpha}(t,0)= -G_{\alpha}(t,0) G_{\bar{\alpha}}(t/0^+,0)
\nonumber \\
  & & +i U \int_C dt_1 G_{\alpha}(t,t_1) G_{\bar{\alpha}}(t/0^+,t_1)
   B_c(t_1,0)\ .
\end{eqnarray}
Finally, we get $B_c^{+-}(t,0)$, $B_c^{--}(t,0)$, and $B_{\alpha}^{+-}(t,0)$ 
from Eq.(4), and upon taking a Fourier transformation,
$B_{\alpha}^{+-}(\epsilon)$ can be expressed in terms of 
$G_{\alpha}^{++}(\epsilon)$, $G_{\alpha}^{+-}(\epsilon)$,
$G_{\alpha}^{-+}(\epsilon)$, and $G_{\alpha}^{--}(\epsilon)$---these 
four Green's functions are directly related to $G^r_{\alpha}$ and
$G^<_{\alpha}$. This completes the analytical derivations.

From Eqs. (2,3,4), the intradot occupation number $n_{\alpha}$ and the 
vertical flow $i_{\alpha}(\epsilon)$ are determined self-consistently.  Let's 
consider a two-probe ($n=L,R$) device and the wideband approximation in which
$\Gamma_n(\epsilon)$ is independent of energy $\epsilon$. Fig.2 plots
the vertical flow $i(\epsilon)=\sum_{\alpha} i_{\alpha}(\epsilon)$ (solid
line) as well as the current density $j_{L/R}(\epsilon)$ 
at a high bias. Clearly, the vertical flow $i(\epsilon)$ is 
non-zero due to the TET processes, and its value $|i(\epsilon)|$ has four 
peaks at energies $\epsilon_{\uparrow}$, $\epsilon_{\downarrow}$, 
$\epsilon_{\uparrow}+U$, and $\epsilon_{\downarrow}+U$, respectively. This 
means that the incident electron indeed can vary its energy in QD by
e-e scattering. In contrast, in a typical SET process the electron keeps 
its energy and does not induce any vertical flow $i(\epsilon)$.

The physics of the TET process that induces the vertical flow is shown
by inset (a) of Fig.2. To start, two incident electrons from left lead
having energies $\epsilon_{\downarrow}$ and $\epsilon_{\uparrow} +U$ tunnel
into the QD. They scatter with each other inside the QD and exchange energy 
to final states $\epsilon_{\uparrow}$ and $\epsilon_{\downarrow} +U$.
Afterwards they tunnel out of the QD. In the vertical flow curve of Fig.2, 
two peaks at $\epsilon_{\downarrow}$ and $\epsilon_{\uparrow}+U$ 
are negative (dips), and the other two peaks at $\epsilon_{\uparrow}$ and
$\epsilon_{\downarrow} +U$ are positive: precisely indicating the 
transfer of states from the initial ones at $\epsilon_{\downarrow}$ and
$\epsilon_{\uparrow} +U$ to the final ones at $\epsilon_{\uparrow}$
and $\epsilon_{\downarrow} +U$. 
It is also worth mentioning that besides the new TET process, 
the usual SET processes also exist in charge transport through the QD
in the present case.

So far we have demonstrated that TET processes can exist. In the following 
we analyze several important questions concerning TET. What is its 
consequence? (i) TET makes current {\it density} a non-conserved quantity,
{\it i.e.} $\sum_n j_n(\epsilon) \not=0$. Of course, the total
current is still conserved, {\it i.e.} $\sum_n I_n =0$. This can be easily
proved from the definition of the vertical flow $i(\epsilon)$, namely
$i(\epsilon)$ has the property $\int i(\epsilon) d\epsilon =0$. 
(ii) TET can involve high energy empty states, namely states which are
higher by about $U$ than the highest chemical potential $max(\mu_L, \mu_R)$
(see inset (a) of Fig.2); it may also involve electrons deep inside the
Fermi sea, namely states which are lower by about $U$ than the lowest chemical 
potential $min(\mu_L, \mu_R)$ (see TET process shown in inset (b) of Fig.2).
(iii) TET may induce a current density that is flowing {\it out} from the 
{\it high} voltage terminal, {\it i.e.} the left lead (indicated by 
the negative peak at $\epsilon_{\downarrow} +U$ in $j_L(\epsilon)$ curve, and
by the arrow A in inset (a) of Fig.2). Similarly, TET may also induce 
a current density that is flowing {\it in} from the low voltage terminal, 
{\it i.e.} the right lead (see arrow A in inset (b) of Fig.2).
These characteristics are rather different from the typical elastic SET 
processes.

Under what conditions does TET or the vertical flow $i(\epsilon)$ exist? 
(i) We found that an increase (decrease) of temperature $T$ or linewidth 
$\Gamma$ will widen (narrow) the peaks of vertical flow, but does not affect 
peak positions and heights significantly. (ii) If $U=0$, the vertical flow 
$i(\epsilon)=0$ identically: TET crucially depends on this parameter.
(iii) If $U\rightarrow \infty$, $i(\epsilon)$ tends to zero. This is because 
at large $U$, the intradot two-electron occupation is prohibited therefore TET
is blockaded. In this case only SET processes occur. (iv) When bias 
potential $eV=\mu_L-\mu_R$ is less than $U$, $i(\epsilon)$ decreases 
drastically. In the limit of $eV=0$, $i(\epsilon)=j_n(\epsilon)=0$. 

Are observable quantities of charge transport affected by the TET process?
(i) Clearly the current density $j_{L/R}(\epsilon)$ is affected significantly
as already discussed above. (ii) In general, the current, conductance, and 
$n_{\alpha}$ will be affected significantly by TET (see below). The current
noise, which reflects the e-e time correlation, will increase due to TET 
processes. However, if one uses the wideband approximation, the TET
dependence in charge current will be lost.

In the rest of this paper, we apply the property of TET to design a
device so that electrons can be pumped from a lead with lower chemical 
potential to another lead having a higher chemical potential. Consider a 
device with three leads ($n=1,2,3$) and consider the non-wideband case. 
We use a model of quasi-square bands where the coupling
$\Gamma_n(\epsilon) =\Gamma /\{ exp[(|\epsilon-c_n|-W)/0.05]+1\}$, 
the width of the band is set by $2W=1$ and its center at $c_n$ which is 
dependent on the terminal voltage $eV_n=\mu_n$ but $\mu_n-c_n$ is kept fixed. 
More specifically, let's assume lead 1 to be a p-type semiconductor with 
$\mu_1-c_1=0.4$; lead 2 an n-type semiconductor with $\mu_2-c_2=-0.4$; and 
lead 3 a metal with $\mu_3-c_3=0$ (see Fig.4). The energy diagram of the 
device is set by external voltages as that shown in Fig.4 so that
$\mu_2>\mu_1>\mu_3$. 
The current density $j_n(\epsilon)$ in this case is 
shown in Fig.3. 
We note that $j_1(\epsilon)$ (dotted line) has two positive peaks at 
$\epsilon_{\downarrow}$ and $\epsilon_{\uparrow} +U$; $j_2(\epsilon)$ and 
$j_3(\epsilon)$ each has one negative peak at $\epsilon_{\downarrow} +U$ 
and $\epsilon_{\uparrow}$, respectively; 
and $i(\epsilon)=-\sum_nj_n(\epsilon)$ has two negative 
peaks at $\epsilon_{\downarrow}$ and $\epsilon_{\uparrow} +U$, and two 
positive peaks at $\epsilon_{\downarrow} +U$ and $\epsilon_{\uparrow}$. 
The current $I_n=\int j_n(\epsilon)d\epsilon$ is quite large.  
We emphasize two points for this pump. (i) In this device the SET process 
almost does not occur because bands of different leads do not overlap. 
Then, clearly, the large current $I_n$ originates from the TET process: 
$I_n\rightarrow 0$ if the vertical flow $i(\epsilon)\rightarrow 0$.
This demonstrates that TET can significantly affect charge current in the
general case of non-wideband coupling. (ii) The charge current in the
terminal with the highest bias voltage, {\it e.g.} lead 2, is {\it negative}
(solid line in the inset of Fig.3), which demonstrates the pump effect.
The pumps works because when an electron tunnels from lead 1 to 3, it 
emits energy $U$ to pump another electron from lead 1 to lead 2, through the 
TET process.

More clearly, the working principle of the TET pump is summarized in Fig.4.
(a) We start from the situation where no charge is in the QD so that levels
$\epsilon_{\uparrow}$ and $\epsilon_{\downarrow}$ are empty.  In this 
situation, an electron in the Fermi sea of lead 1 having energy 
$\epsilon_{\downarrow}$ can easily tunnel into the QD (Fig.4a). 
(b) After this electron tunnels into the QD and occupies the QD level of
$\epsilon_{\downarrow}$, the other intradot level $\epsilon_{\uparrow}$ 
is raised up to $\epsilon_{\uparrow}+U$, so that another electron with 
energy $\epsilon_{\uparrow}+U$ in lead 1 tunnels into the QD (Fig.4b).
(c) When the second electron comes into QD, due to the e-e Coulomb interaction
$U$, the level $\epsilon_{\downarrow}$ with its electron is raised up to
$\epsilon_{\downarrow}+U$, leading to a negative peak at
$\epsilon_{\downarrow}$ and a positive peak at $\epsilon_{\downarrow}+U$
in the vertical flow curve $i(\epsilon)$. Now the intradot two-electron 
system has total energy $\epsilon_{\uparrow}+\epsilon_{\downarrow}+U$ 
(Fig.4c). Afterwards the first electron in state $\epsilon_{\downarrow}+U$
easily tunnels to lead 2 and takes away energy $\epsilon_{\downarrow}+U$.
The net effect is that the two electrons exchanged energy $U$, which is the 
TET process discussed above. When the first electron leaves the QD,
the other electron at $\epsilon_{\uparrow}+U$ falls down to
$\epsilon_{\uparrow}$, leading to a negative peak at $\epsilon_{\uparrow}+U$ 
and a positive peak at $\epsilon_{\uparrow}$ in the curve of $i(\epsilon)$.
(d) Finally, the second electron tunnels to lead 3 (Fig.4d) and our device
returns to its initial conformation of Fig.4a. This way an electron is pumped 
from lead 1 to lead 2, where $\mu_1<\mu_2$, via the TET process.
We emphasize that each tunneling event from Fig.4a-d is a first-order normal
tunneling event in which tunneling occurs at two aligning states
(not like higher-order virtual co-tunneling process)\cite{ref11}. We
therefore conclude that the TET process should have large probability to occur
so that $i(\epsilon)$ can be near the unit value $e/h$ (see Fig.3).
We have also investigated the terminal voltage ({\it e.g.} $V_3$) 
dependence of current $I_n$, shown in the inset of Fig.3. As $V_3$ is 
increased so that $eV_3=\mu_3$ passes the lowest resonance 
state $\epsilon_{\uparrow}$, the tunneling event in Fig.4d can not occur 
and the TET process is blockaded, leading to a significant reduction 
of all currents $I_n$ (including the pumping current $I_2$) (see inset of
Fig.3). 

In summary, we have investigated the two-electron correlated scattering 
process in mesoscopic system. TET induces a vertical flow in 
the scattering region so that electrons enter and exit the device with 
different energies. TET is found to affect current density significantly, 
and the process can involve high empty states or/and low filled states of
the leads. The properties of TET suggests an interesting working principle 
of an electron pump which pumps charge carries from a lead with low chemical 
potential to another lead with a higher chemical potential. In fact, if the 
bands of lead 1 is full and bands of leads 2 and 3 are empty, these results 
are not affected. Our proposed pump is very different as compared to the 
electron-photon or parametric pumps\cite{ref9,ref10}. It should be 
experimentally feasible even for devices fabricated in two-dimensional 
electron gas. In that case, the bands in our theory can be replaced with
Landau levels.

{\bf Acknowledgments:} We gratefully acknowledge financial support from
NSERC of Canada, FCAR of Quebec (Q.S., H.G), and a RGC grant from the
SAR Government of Hong Kong under grant number HKU 7091/01P (J.W.).


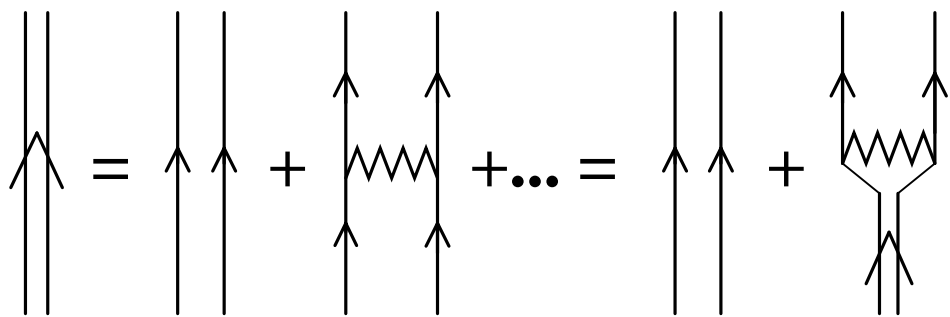
\begin{figure}
\caption{
Relevant Feynman diagrams considered in this work. The single solid lines,
doubles solid lines, and the wave lines stand for the Green's functions
$G_{\alpha}$, $B_{\alpha/c}$, and interaction $U$, respectively.
}
\label{fig1}
\end{figure}

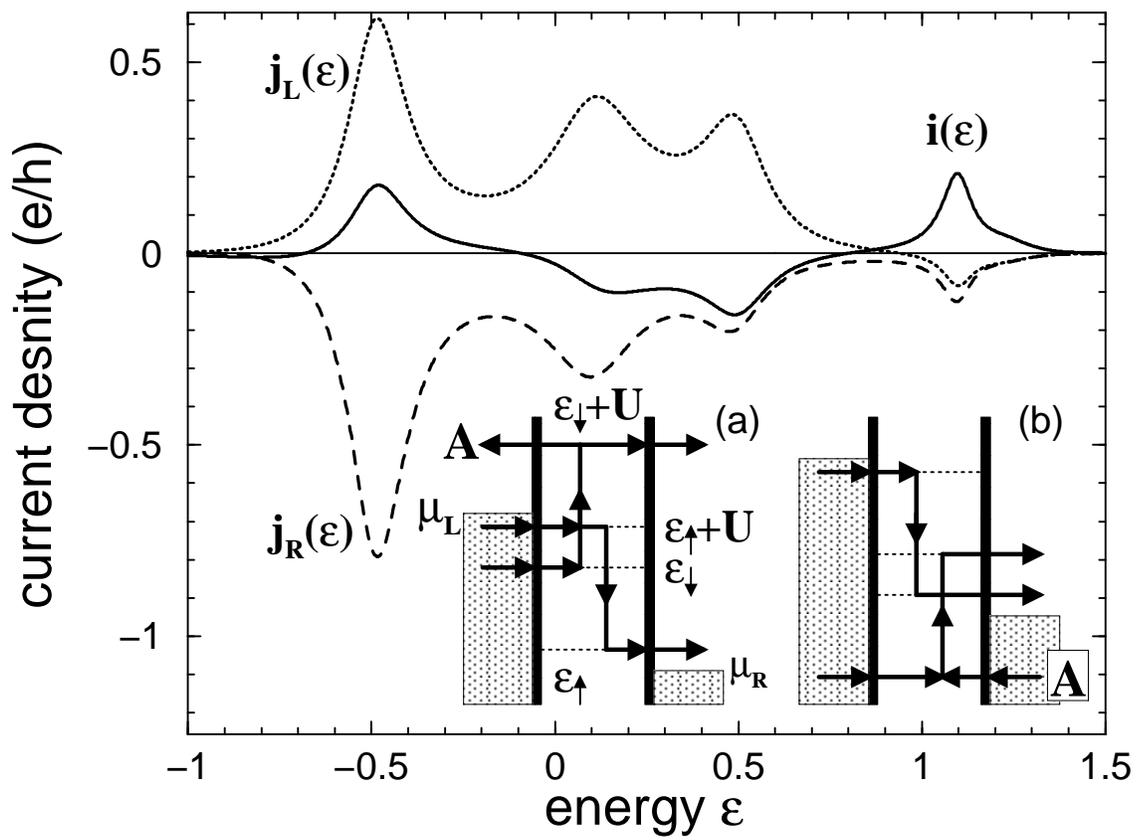
\begin{figure} 
\caption{ 
$j_{L/R}(\epsilon)$ and $i(\epsilon)$ versus energy $\epsilon$ in 
the wideband limit. The parameters are: $\epsilon_{\uparrow}=-0.5$, 
$\epsilon_{\downarrow}=0.1$, $U=1$, $T=\Gamma=0.1$, and $\mu_L=-\mu_R=0.7$. 
Insets (a) and (b) are schematic plots for two kinds of TET processes.
} 
\label{fig2}
\end{figure}

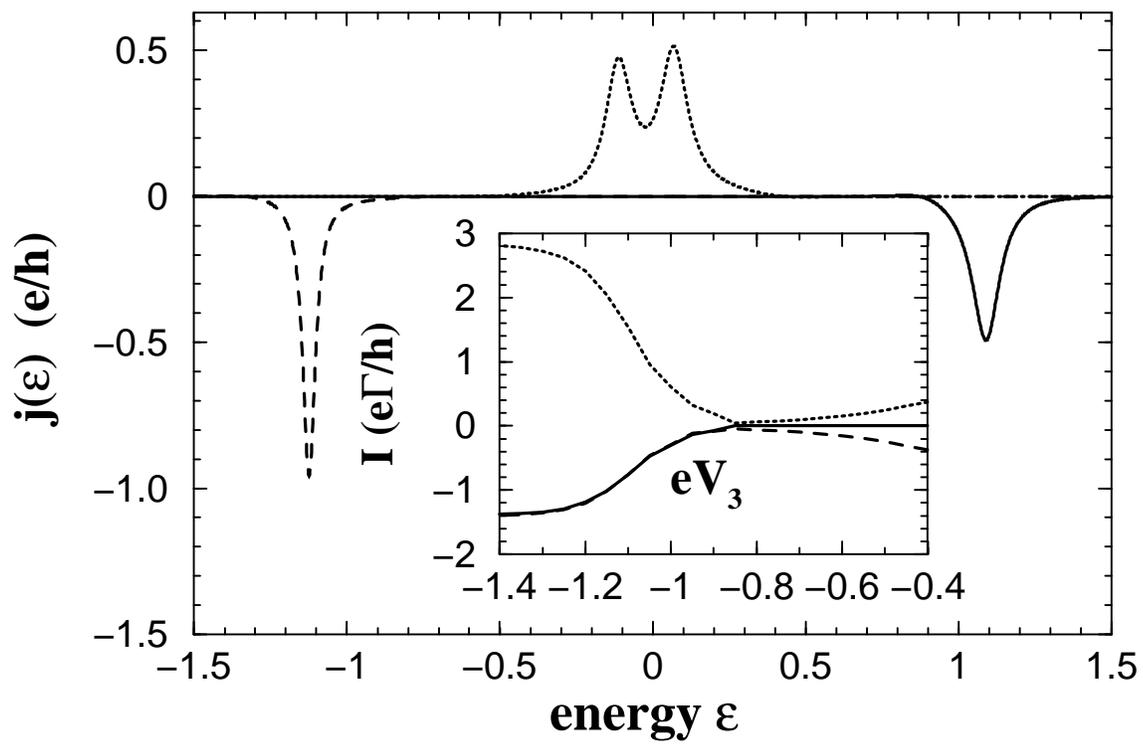
\begin{figure} 
\caption{ 
The main plot shows $j_n(\epsilon)$ versus $\epsilon$ with $\mu_3=-1.3$ and 
the inset shows $I_n$ versus $eV_3=\mu_3$ for the non-wideband case. 
Other parameters are: $\epsilon_{\uparrow}=-1.1$, 
$\epsilon_{\downarrow}=-0.1$, $U=1.2$, $T=\Gamma=0.05$,
$\mu_1=0.4$, and $\mu_2=0.9$. The dotted, solid, dashed curves in the
main plot correspond to $j_1(\epsilon)$, $j_2(\epsilon)$, and 
$j_3(\epsilon)$, respectively; they correspond to $I_1,I_2,I_3$ in the
inset.
} 
\label{fig3}
\end{figure}

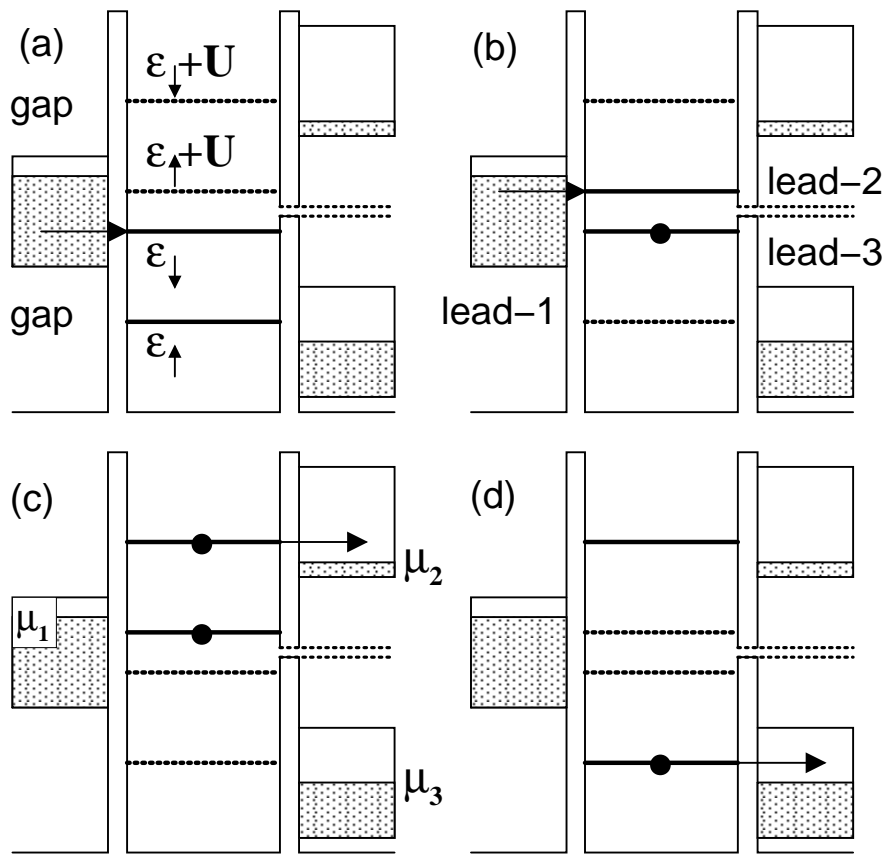
\begin{figure} 
\caption{ 
Schematic plots for the working principle of the TET charge pump.
} 
\label{fig4}
\end{figure}


\begin{references} 

\bibitem{datta_book}  
S. Datta, {\sl electronic transport in mesoscopic systems} (Cambridge Univ.
Press 1995); H. Haug and A.-P. Jauho, {\sl Quantum Kinetics in Transport and
Optics of Semi-conductors} (Springer-Verlag 1998).

\bibitem{meir1}
Y. Meir and N.S. Wingreen, Phys. Rev. Lett. {\bf 68}, 2512 (1992).

\bibitem{ref4}
Q.-f. Sun and H. Guo, Phys. Rev. B {\bf 64}, 153306 (2001).

\bibitem{ref5}
Y. Meir, N.S. Wingreen, and P.A. Lee, Phys. Rev. Lett. {\bf 66}, 3048 (1991);
{\sl ibid} {\bf 70}, 2601 (1993).

\bibitem{ref6}
T.-K. Ng, Phys. Rev. Lett. {\bf 76}, 487 (1996).

\bibitem{ref7}
N.S. Wingreen and Y. Meir Phys. Rev. B {\bf 49}, 11040 (1994).

\bibitem{ref11}
S. De Franceschi, {\sl et al}. Phys. Rev. Lett. {\bf 86}, 878 (2001).

\bibitem{ref9} 
C.A. Stafford and N.S. Wingreen, Phys. Rev. Lett. {\bf 76}, 1916 (1996); 
T.H. Oosterkamp, {\sl et al}. Nature {\bf 395}, 873 (1998).

\bibitem{ref10} 
Y. Wei, J. Wang, and H. Guo, Phys. Rev. B {\bf 62}, 9947 (2000); 
P. Sharma and C. Chamon, Phys. Rev. Lett. {\bf 87}, 096401 (2001).


\end{references}
\end{document}